\documentclass[preprint,review,12pt,time]{elsarticle}
\pdfoutput=1

\usepackage{amssymb}
\usepackage{amsmath}
\usepackage{float}
\usepackage{setspace}
\usepackage{hyperref}
\hypersetup{colorlinks=true, linkcolor=blue, anchorcolor=blue, citecolor=blue}
 \usepackage{lineno}

\begin{document}

\begin{frontmatter}



\title{Multi-objective optimization of longitudinal injection based on a multi-frequency RF system for fourth-generation storage ring-based light sources}


\author[1,2]{Weihang Liu}

\address[1]{Institute of High Energy Physics, Chinese Academy of Sciences, Beijing 100049, China}

\author[1]{Yi Jiao\corref{cor1}}

\ead{jiaoyi@ihep.ac.cn}

\author[1,2]{Yu Zhao}
\author[1,2]{Jianliang Chen}
\author[1,2]{Yanliang Han}
\author[1,2]{Sheng Wang\corref{cor1}}
\ead{wangs@ihep.ac.cn}
\address[2]{Spallation Neutron Source Science Center, Dongguan 523803, China}

\cortext[cor1]{Corresponding author}

\begin{abstract}
In the fourth-generation storage ring light sources (4GLSs), associated with the extremely strong nonlinearities inherent in the multi-bend achromat design, the dynamic acceptance is usually small and it is difficult to implement traditional off-axis local-bump injection. To release the requirement on dynamic acceptance, on-axis longitudinal injection schemes have been explored. In this paper, we present a multi-objective optimization of longitudinal injection with a multi-frequency RF system, based on the parameters of the Southern Advanced Photon Source, a 4GLS proposed in China. We show that by treating the optimal bunch lengthening condition as an optimizing objective rather than a condition that must be satisfied, a plethora of feasible candidate solutions can be found, showing different trade-offs among multi-objectives. From these candidate solutions, one can find an optimal RF parameter setting that is most adapted to a specific 4GLS physics design and the available technical level of injection kicker. Especially, it would be feasible to realize longitudinal injection within a static bucket enabled by a double-frequency RF system, suggesting an attractive longitudinal injection option for 4GLSs.

\end{abstract}


\end{frontmatter}


\section{Introduction}\label{sec.I}
The fourth-generation storage ring-based light sources (4GLSs) \cite{hettel_dlsr_2014} are capable of producing ultralow-emittance electron beams and (quasi-) diffraction-limited electron beams, and hence pushing the brightness and coherence of X-rays well beyond existing third-generation light sources (3GLSs). To reach an ultralow emittance within a reasonable circumference, it usually adopts a multi-bend achromat (MBA) \cite{einfeld_first_2014} lattice with a compact layout and strong quadrupoles in the storage ring design of a 4GLS. The MBA lattice intrinsically has small dispersions and large negative chromaticities. The chromaticities should be corrected to be above zero to mitigate the collective instabilities and avoid large tune shifts with energy deviation. To achieve this, strong achromatic sextupoles are adopted, which, however, introduce large nonlinearities to the beam dynamics of the ring. Accordingly, the dynamic aperture (DA) of the storage ring of a 4GLS is noticeably reduced compared with that available for most 3GLSs. This is especially true when pushing the emittance down to be close to its lowest limit to achieve the highest possible brightness. Consequently, the traditional off-axis local-bump injection, which is widely adopted in 3GLSs and requires DA of order 10 mm, is not desirable for a 4GLS. Thus, other injection schemes with released DA requirements have been extensively studied, such as pulsed multipole injection (\cite{takaki_beam_2010,jiao_pulsed_2013}, off-axis injection as well but requires smaller DA, saying ~5 mm) and novel on-axis injection schemes \cite{xiao_-axis_2013,aiba_longitudinal_2015,jiang_using_2016,xu_-axis_2016,tordeux_longitudinal_2017,jiang_-axis_2018,kim_injection_2019}.

In the so-called on-axis ‘swap-out’ injection scheme \cite{xiao_-axis_2013}, a fast kicker is used to knock out one or more stored bunches and replace it or them with fresh bunches coming from the injector. This scheme only requires a DA sufficiently larger than a few times the injected beam size and has been adopted by several 4GLSs, such as APS-U, HEPS and ALS-U \cite{steier_-axis_2017,borland_upgrade_2018,jiao_heps_2018}. To avoid the stored bunches being disturbed by the kicker, the pulse length of the kicker should be short enough, e.g., smaller than two times of the length of a bucket (the region of stable or bounded motions of a particle in longitudinal phase space) and on the scale of nanoseconds. Note that this scheme does not allow accumulation and requires an injector capable of generating full-charge bunches with high stability. 

Another on-axis injection option, longitudinal injection, is being explored. Several longitudinal injection schemes \cite{aiba_longitudinal_2015,jiang_using_2016,xu_-axis_2016,tordeux_longitudinal_2017,jiang_-axis_2018} that promise accumulation have been proposed, where the bunch is injected on-axis transversely but with an energy and/or phase deviation relative to the synchronous particle. The principles of these schemes are similar. Taking the first proposed longitudinal injection scheme \cite{aiba_longitudinal_2015} as an example, a bunch of off-energy and off-phase is injected into a bucket  and then gradually merges with the stored bunch of the same bucket through radiation damping effect. 

Compared with the swap-out injection, longitudinal injection also requires sufficient off-energy dynamics acceptance, to accommodate large-amplitude synchrotron oscillations of the injected bunch. Also, this scheme generally imposes a more stringent requirement on the kicker pulse length than for swap-out injection, based on the consideration that the falling time of the kicker pulse should be smaller than the separation between the injected and the stored bunches in the same bucket to avoid the stored bunch being disturbed by the kicker. On the other hand, the available minimum falling time of the kicker pulse is limited to about 2 ns at the current state of the art \cite{naito_multibunch_2011,nakamura_bucket-by-bucket_2011,chen_strip-line_2019,wang_novel_2021}. Thus, it is important for longitudinal injection to have a released requirement on the kicker pulse as much as possible.

The first proposed longitudinal injection scheme \cite{aiba_longitudinal_2015} considered a single-frequency, 100 MHz RF system. Later proposals \cite{jiang_using_2016,xu_-axis_2016} focused on the case with a double-frequency RF system, where a RF system with a higher fundamental RF frequency can be allowed. To achieve a time separation between injected and stored bunches of longer than 2 ns, RF gymnastics is proposed by appropriately altering the voltages and phases of RF cavities during the injection. However, in such a scheme, the bucket is not static and requires frequent changes in the RF parameters over a wide range. Consequently, the bunch length of the stored bunch may decrease dramatically during injection, leading to strong collective effects and even beam loss. More efforts on both physics studies and RF experiments are necessary to ensure that the frequent variation of RF parameters does not cause any risk to the stable operation of the RF system and the machine. 

Apparently, invariant RF parameters and a static bucket during injection are more desirable. Further study \cite{jiang_-axis_2018} showed that with a triple-frequency RF system, it is feasible to simultaneously realize static RF parameters and sufficient time separation between injected and stored bunches. In this study \cite{jiang_-axis_2018}, the required RF parameters were derived under the optimal lengthening condition for the stored bunch. 

However, we note that it would be more common for an actual machine to be operated at a state not far from, rather than exactly at the optimal lengthening condition. If the optimal bunch lengthening is not a condition that must be satisfied, more free variables will be available, and more candidate solutions may be found. From these solutions, one can select one that adapts well to a given light source design. Following this idea, we implement a multi-objective optimization study for longitudinal injection with a multi-frequency RF system. By taking the Southern Advanced Photon Source (SAPS) \cite{wang_proposal_2021,zhao_design_2021}, a 4GLS proposed in China, as an example, we demonstrate that with a double-frequency RF system, it is feasible to realize longitudinal injection with a static bucket and sufficient time separation; and if with more frequencies of the RF system, it is feasible to increase the time separation between injected and stored bunch to the best value, i.e., half of the separation of two adjacent buckets.

The rest of the paper is arranged as follows. The idea of multi-objective optimization of longitudinal injection is described in Sec. 2. And in Sec. 3, optimization designs of longitudinal injection with a double-, triple- and quadruple-frequency RF system are performed based on the SAPS parameters. Conclusions are given in Sec. 4. 

\section{Longitudinal injection with a multi-frequency RF system}\label{sec.2}
 The longitudinal motion of a particle in a storage ring, subject to RF cavities, is determined by the Hamiltonian \cite{chao_lectures_2020}
\begin{equation}\label{eq1}
H=\frac{1}{2} \omega_{0} h \alpha_{c} \delta^{2}+\frac{U_{0}}{E_{0} T_{0}} \Phi,
\end{equation}
where $\omega_0$ is the angular revolution frequency, $h$ is the bucket number, $\alpha_c$ is the momentum compaction factor, $\delta$ is the relative energy deviation, $U_0$ is energy loss per turn, $E_0$ is the nominal beam energy, $T_0$ is the revolution time of the ring, and $\Phi$ is the normalized potential of the $N$ frequencies RF system, which has the form (see Appendix for details) 
\begin{equation}\label{eq2}
\Phi=\phi+\sum_{i=1}^{N} \frac{v_{i}}{n_{i}}\left[\cos \left(n_{j} \phi+\phi_{i}\right)-\cos \left(\phi_{i}\right)\right],
\end{equation}
where $\phi$ is the RF phase of a particle, $v_i$ is the normalized voltage of the $i^{th}$ frequency cavity, in unit of $U_0$; $n_i$ is the harmonic number of the $i^{th}$ frequency ($n_1$ corresponds to the fundamental frequency and $n_1\equiv 1$), $\phi_i$ is the phase of the synchronous particle in the $i^{th}$ frequency cavity.

\begin{figure}[H]
	\centering
	\includegraphics[width=10cm]{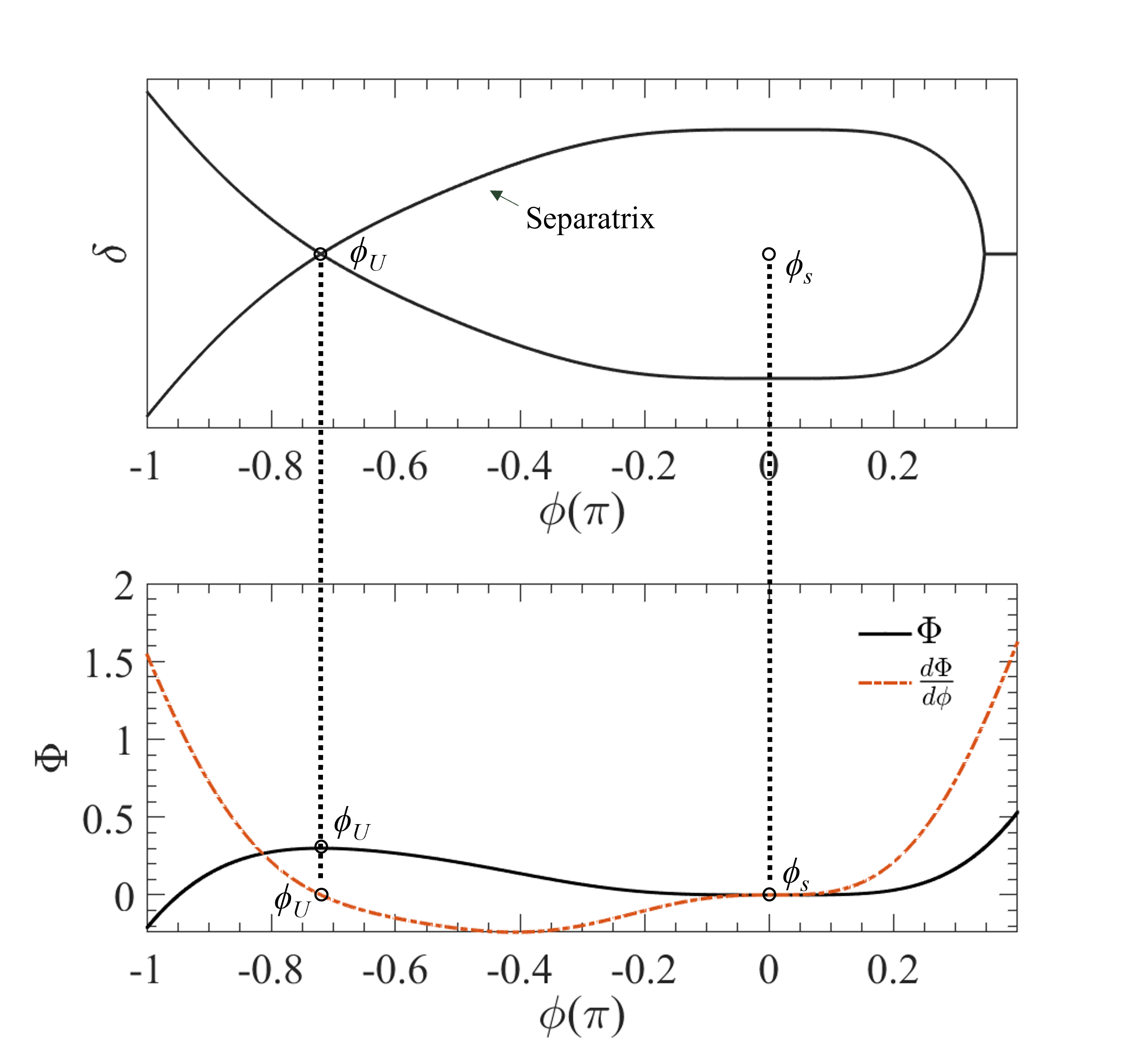}
	\caption{Diagram of a bucket (upper) and the corresponding potential function (lower) subject to a triple-frequency RF system with harmonic number of (1, 2, 3). The subscript $s$ for $\phi$ stands for stable, and $U$ for $\phi$ stands for unstable. The parameters used to show this diagram are $v_1 = -1.767$, $v_2 = -0.789$, $v_3= 0.225$, $\phi_1 = -0.6$, $\phi_2 = -2.851$ and $\phi_3 = -1.651$.}
	\label{Fig1}
\end{figure}

Figure \ref{Fig1} shows a bucket and the corresponding normalized potential for illustration. In a bucket, there are two fixed points. One fixed point is stable (with phase of $\phi_s = 0$) and the other is unstable (with phase of $\phi_U$). Both fixed points are at the axis $\delta = 0$. The phases of these two fixed points can be obtained by 
\begin{equation}\label{eq3}
\frac{d \Phi}{d \phi}|_{0}=0,\frac{d \Phi}{d \phi}|_{\phi_{U}}=0.
\end{equation}

To ensure the motions starting from the points between these two fixed points are bounded, the normalized potential $\Phi$ should satisfy 
\begin{equation}\label{eq4}
\Phi|_{\phi}>0,\frac{d \Phi}{d \phi}|_{\phi}<0, \phi \in\left(\phi_{U}, 0\right). 
\end{equation}

The stored bunch center is at the stable fixed point. For longitudinal injection in such a bucket, the injected bunch (on-energy, off-phase) is considered to be injected close to the unstable fixed point and inside the bucket, so as to obtain a large separation between injected and stored bunch. For simplicity, here we use the separation between two fixed points, $|\phi_U|$, as an indicator of the separation between injected and stored bunch. It needs to mention here that the latter should be somewhat smaller than the former in practice.

Considering that the separation between two adjacent buckets is $2\pi$, the optimal value of $|\phi_U|$ is about $\pi$, corresponding to a largest possible separation between injected and stored bunch. Further increasing $|\phi_U|$ will relax the requirement of the falling time but strengthen the requirement of the rising time of the kicker pulse.

In addition, considering the technical level of the kicker pulse (the available minimum falling time is $\sim 2$ ns), $|\phi_U|$ should has a sufficiently large value (if it is difficult for $|\phi_U|$ to be close to $\pi$). The phase separation $|\phi_U|$ is related to the time separation $\Delta t$ by 
\begin{equation}
\Delta t=\frac{\left|\phi_{U}\right|}{h \omega_{0}}=\frac{\left|\phi_{U}\right|}{2 \pi f_{0}},
\end{equation}
where $f_0$ is the fundamental frequency of the RF system. The technical limit $\Delta t \sim 2$ ns, imposes a lower limit for the value of $|\phi_U|$, i.e., $2\,{\rm ns}\times f_0\times 2\pi$. 

To ensure existence of feasible solutions for longitudinal injection, the fundament frequency should be low enough such that $2\,{\rm ns}\times f_0\times 2\pi \textless \pi $. And in the design of longitudinal injection, one important goal is to choose the RF parameters such that $|\phi_U|$ is sufficiently larger than $2\,{\rm ns}\times f_0\times 2\pi$ and is best to be close to $\pi$.

On the other hand, the bucket height determined by the chosen RF parameters, $\delta_m$, should not be larger than the MA determined by transverse dynamics \cite{steier_measuring_2002}. The bucket height is of the form
\begin{equation}\label{eq6}
\delta_{m}=\sqrt{\frac{U_{0}}{\pi h \alpha_{c} E_{0}} \Phi\left(\phi_{U}\right)}.
\end{equation}

In a 4GLS, due to strong nonlinearities by strong sextupoles, even with thorough optimization, the available MA determined by transverse dynamics is typically 3\% $\sim$ 4\% \cite{jiao_statistical_2017}. If $\delta_m$ is larger than the MA, the injected bunch will have a large-amplitude synchrotron motion and get lost at a certain number of revolutions after injected into the ring. 

Also, if using a multi-frequency RF system, in a 4GLS, it is common and desirable to stretch the beam length of the stored bunch to mitigate various collective effects \cite{nagaoka_collective_2014}. An optimal bunch lengthening state will be reached when the following condition is satisfied,
\begin{equation}\label{eq7}
\frac{d^{2} \Phi}{d \phi^{2}}|_{0}=0.
\end{equation}
However, it is thought Eq.\eqref{eq7} is not a necessary condition that must be satisfied for either longitudinal injection or daily operation of the ring, and it is enough to make the absolute value of the second derivative of $\Phi$ to be close to zero. 

In brief, there are mainly three optimizing objectives, $|\phi_U|$, $\delta_m$ and $|\frac{d^2\Phi(0)}{d^2\phi^2}|$, in the design of longitudinal injection with a multi-frequency RF system. It is possible to simultaneously optimize these three objectives using a multi-objective optimization algorithm. Nevertheless, in the presented optimization, we make a two-objective optimization with the algorithm of multi-objective particle swarm optimization (MOPSO) \cite{kennedy_particle_1995,clerc_particle_2002}, which has been applied to many accelerator optimization problems \cite{zhao_design_2021,huang_nonlinear_2014,huang_online_2015,jiao_optimizing_2017}. In the opinion of authors, for longitudinal injection design  $|\phi_U|$ (bunch separation) is somewhat more important than the other two objectives. Thus we choose to optimize $\delta_m$ and $|\frac{d^2\Phi(0)}{d^2\phi^2}|$ for a specific value of $|\phi_U|$. As will be shown in the next section, doing in this way allows fast convergence, and especially, helps to investigate the capability of different RF frequency combinations to increase the separation between injected and stored bunch.

Here we express the optimization in the following form
\begin{equation}\label{eq8}
\begin{array}{ll}
\min & {\left[\delta_{m}\left(v_{1}, \phi_{1}, \ldots, v_{N}, \phi_{N}\right),\left|\frac{d^{2} \Phi}{d \phi^{2}}|_{0}\left(v_{1}, \phi_{1}, \ldots, v_{N}, \phi_{N}\right)\right| \right]} \\
\text { subject to } &\frac{d \Phi}{d \phi}|_{0}=0, \\
& \frac{d \Phi}{d \phi}|_{\phi_{U}}=0, \\
& \Phi|_{\phi}>0, \quad \phi \in\left(\phi_{U}, 0\right), \\
& \frac{d \Phi}{d \phi}|_{\phi}<0, \quad \phi \in\left(\phi_{U}, 0\right).
\end{array}
\end{equation}

\section{Numerical demonstrations with parameters of the SAPS}\label{sec.3}

SAPS \cite{wang_proposal_2021,zhao_design_2021} is a 4GLS proposed to be built in Guangdong province of China, adjacent to the China Spallation Neutron Source \cite{wei_china_2009,sheng_introduction_2009}. After a few years of study, a baseline design for the SAPS storage ring has been proposed \cite{zhao_design_2021}, with a hybrid seven-bend-achromat lattice and natural emittance of about 23 pm at 3.5 GeV. Main parameters of the SAPS storage ring are listed in Table 1. As a candidate option for the injector, a C-band full energy linac is under consideration \cite{liu_preliminary_2021}, with beam parameters listed in Table 2. The bunch charge of the injected bunch provided by the linac is less than the design bunch charge of the SAPS storage ring (100 pC vs. 3300 pC). Accordingly, for this injector option accumulation injection is required and longitudinal injection is considered.

In the following, we take SAPS as an example to demonstrate the proposed multi-objective optimization for longitudinal injection with a multi-frequency RF system.

\begin{table}[h]
	\centering
	\caption{Main parameters of the SAPS storage ring}\label{tb1}
	{	\begin{tabular}[h]{lc}
			\\
			\hline
			Parameters          & Value\\
			\hline
			Ring circumference (m)   &972\\
			Beam energy (GeV)        &3.5 \\
			Energy spread (\%)          & 0.12 \\
			Energy loss per turn (MeV)          &0.9  \\
			Fundamental RF cavity frequency (MHz)       & 166.6  \\
			Momentum compaction             & $2.49 \times 10^{-5}$   \\
             Momentum acceptance (\%)       & 4 \\
			\hline
	\end{tabular}}
\end{table}

\begin{table}[h]
	\centering
	\caption{Beam parameters of the full energy linac}\label{tb2}
	{	\begin{tabular}[h]{lc}
			\\
			\hline
			Parameters          & Value\\
			\hline
			Charge (pC)   &100\\
			Nor. Emittance ($\mu$m)        &2 \\
			Bunch length (mm)          & 1.5 \\
			Energy spread (\%)          &0.12  \\
			\hline
	\end{tabular}}
\end{table}

For the SAPS storage ring RF system, the fundamental frequency has been fixed to 166.6 MHz, while the parameters of the harmonic cavities are not determined yet. Next, we will consider longitudinal injection with a double-, triple- and quadruple-frequency RF system, with harmonic numbers as (1, 2), (1, 2, 3), and (1, 2, 3, 4), respectively. In this study $f_0 =$ 166.6 MHz, the lower limit for the value of $|\phi_U|$ is  $2\,{\rm ns}\times 166.6{\rm MHz}\times 2\pi = 0.67\pi $. Here we somewhat arbitrarily choose 0.72$\pi$, a value slightly larger than 0.67$\pi$, as the minimum value of $|\phi_U|$, to leave some space for the injected bunch. Based on similar considerations, we set 1.05$\pi$ as the maximal value of $|\phi_U|$. 

\subsection{Double-frequency RF system}\label{sec.3.1}
With $n_1 = 1$, $n_2 = 2$, the potential of a double frequency RF system is
\begin{equation}\label{eq9}
\Phi=\phi+v_{1}\left[\cos \left(\phi+\phi_{1}\right)-\cos \left(\phi_{1}\right)\right]+\frac{v_{2}}{2}\left[\cos \left(2 \phi+\phi_{2}\right)-\cos \left(\phi_{2}\right)\right].
\end{equation}

For given $\phi_U$, there are four variables in this potential: RF voltages $v_1$, $v_2$ and phases $\phi_1$, $\phi_2$. We use $\phi_1$ and $\phi_2$ as the optimizing variables. For each set of $(\phi_1, \phi_2)$, the voltages $v_1$ and $v_2$ have fixed values, and can be derived from Eq. \eqref{eq3}
\begin{equation}\label{eq10}
v_{1}=\frac{\sin \left(\phi_{2}\right)-\sin \left(\phi_{2}+2 \phi_{U}\right)}{\sin \left(\phi_{2}\right) \sin \left(\phi_{1}+\phi_{U}\right)-\sin \left(\phi_{1}\right) \sin \left(\phi_{2}+2 \phi_{U}\right)},
\end{equation}
\begin{equation}\label{eq11}
v_{2}=\frac{\sin \left(\phi_{1}\right)-\sin \left(\phi_{1}+\phi_{U}\right)}{\sin \left(\phi_{1}\right) \sin \left(\phi_{2}+2 \phi_{U}\right)-\sin \left(\phi_{2}\right) \sin \left(\phi_{1}+\phi_{U}\right)}.
\end{equation}

We setup up the initial population for MOPSO as follows: (a) randomly generate two values in $[-\pi,\pi]$ for $\phi_1$ and $\phi_2$. (b) evaluate the voltages $v_1$ and $v_2$ based on Eqs. \eqref{eq10} and \eqref{eq11}, calculate the values of $\Phi$ and $\frac{d\Phi}{d\phi}$ in $(\phi_U, 0)$. (c) test if the conditions in Eq. \eqref{eq4} are satisfied. If true, save $\phi_1$ and $\phi_2$ as one initial seed. If false, go back to (a). Repeat the process until finding a sufficient number of initial seeds (here this number is 100).

\begin{figure}[H]
	\centering
	\includegraphics[width=8cm]{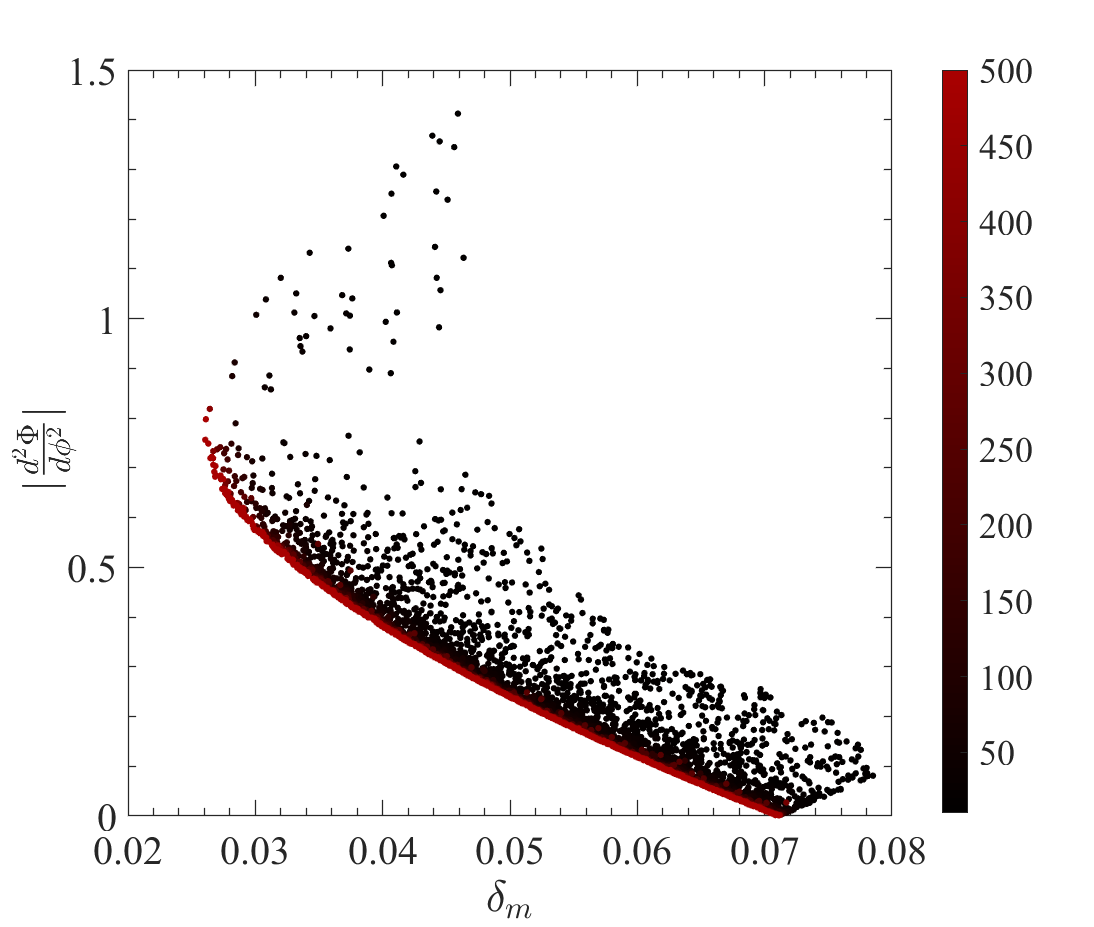}
	\caption{Optimized results in the objective space for longitudinal injection with a double-frequency RF system and with $|\phi_U|=0.72\pi$. The color bar is used to indicate different generations.}
	\label{Fig2}
\end{figure}

MOPSO optimizations are then performed for the case of $|\phi_U|=0.72\pi$, with the evolution of solutions in the objective space shown in Fig.\ref{Fig2}. One can clearly see the convergence of solutions over 400 generations.

The final front solutions show a negative correlation between $\delta_m$ and $|\frac{d^2\Phi(0)}{d^2\phi^2}|$. It suggests a solution satisfying the optimal bunch lengthening condition, i.e., $\frac{d^2\Phi(0)}{d^2\phi^2} = 0$. However, it corresponds to a $\delta_m$ of about 7\%, which is 1.75 times larger than the MA (4\%, see Table 1) determined by the transverse dynamics of the baseline design of the SAPS. Thus, this optimal stretching solution is not feasible for practical longitudinal injection. 

An RF parameter setting with optimal bunch lengthening and $\delta_m$ of 0.04 can be obtained by numerically solving Eqs. \eqref{eq3}, \eqref{eq6}, and \eqref{eq7}. This setting has $|\phi_U|$ of 0.59$\pi$, corresponding to a time separation of 1.77 ns, shorter than 2ns. It is also not feasible for longitudinal injection. This can also be inferred from Fig.\eqref{Fig2}, where this parameter setting corresponds to the point (0.04, 0), well below the final front solutions. 

\begin{figure}[H]
	\centering
	\includegraphics[width=12.5cm]{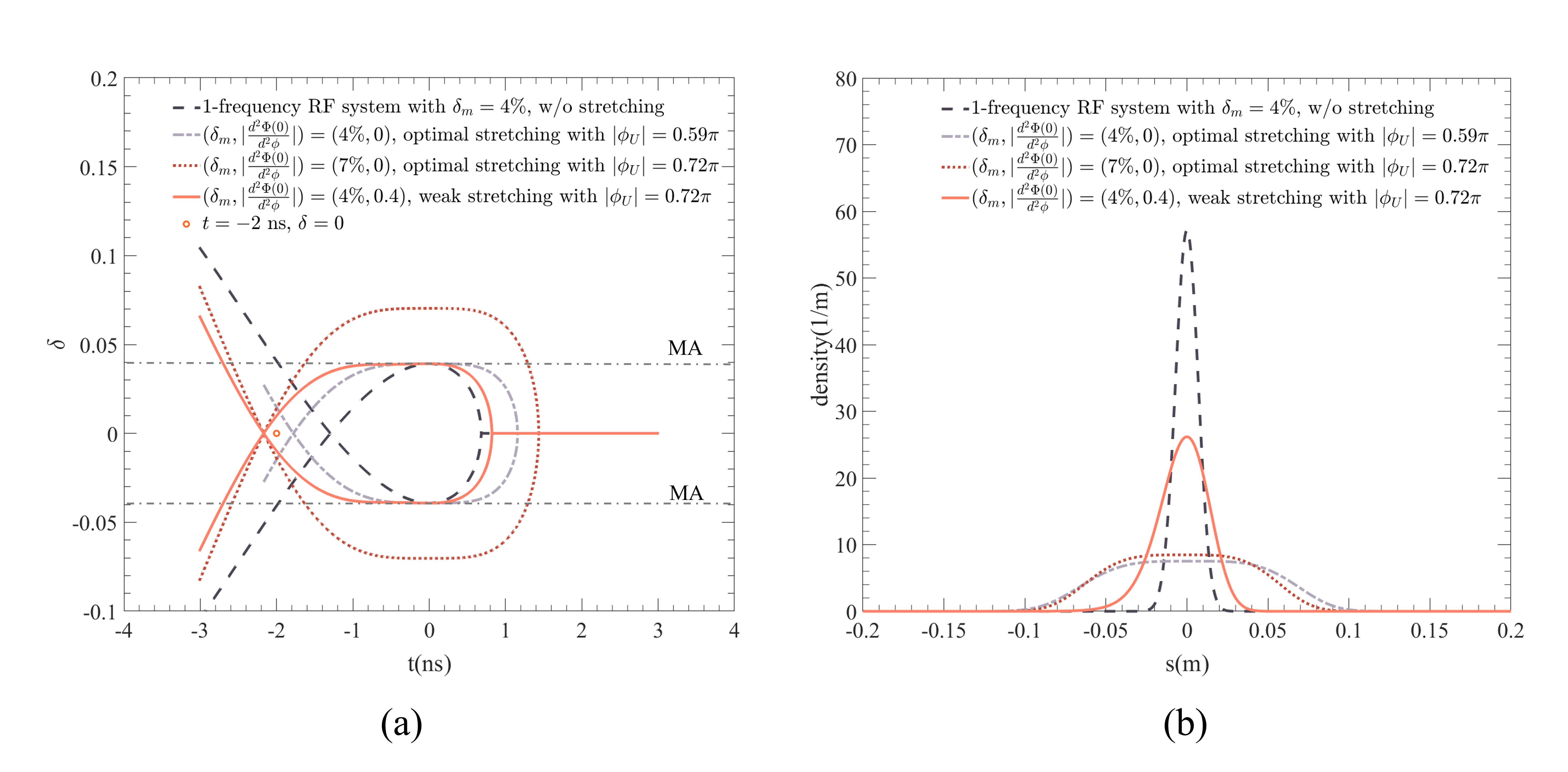}
	\caption{Separatrix (a) and charge density (b) with a double-frequency RF system and different RF parameter settings. The separatrix and charge density in presence of a single-frequency RF system is also presented. The circle denotes the initial position of the injected bunch.}
	\label{Fig3}
\end{figure}

On the other hand, from the final front solutions, one can find a solution (denoted as solution A) with $\delta_m =0.04$ and $|\frac{d^2\Phi(0)}{d^2\phi^2}|=0.4$. Apparently, this solution suggests a weak stretching of the bunch length. Further calculation (see also Fig.\ref{Fig3} and Table 3) indicates that with solution A, the bunch lengthening factor is 2.2, relative to the natural bunch length with only the fundamental-frequency RF cavity. This lengthening factor is comparably smaller than that for optimal stretching, which is about 6, implying stronger collective effects and lower beam lifetime. And consequently, a more frequent injection will be necessary to keep an approximately constant beam current. Nevertheless, this solution does suggest a feasible way for the SAPS to realize longitudinal injection in a static bucket by using a double-frequency RF system.

To verify the feasibility of solution A, we simulate the evolution of the injected bunch after injection with the ELEGANT program \cite{borland_elegant_2000}, based on the SAPS lattice, beam parameters shown in Table 2, and detailed RF parameters listed in Table 3. As shown in Fig. 4, the injected bunch starts from $t = -2$ ns and $\delta = 0$, oscillates in longitudinal phase space, and finally centers itself at the stable fixed point due to radiation damping effect, during which no particle loss is recorded. 

Other cases with $|\phi_U|$ of larger than 0.72$\pi$ are also tested. The final front solutions in the objective space are shown in Fig. 5. It appears that for larger $|\phi_U|$, e.g., $0.85\pi$, in all found solutions, the corresponding bucket heights are larger than 0.04, and there are not feasible solutions for SAPS to implement longitudinal injection with a double-frequency RF system.

\begin{table}[h]
 \centering
\caption{RF voltages, phases and equilibrium beam lengths with different RF parameter settings}\label{tb3}
\resizebox{\textwidth}{30mm}{
\setlength{\tabcolsep}{8mm}{

\begin{tabular}{lcc} 
\hline
\multicolumn{3}{c}{single-frequency
  RF system with bucket height of 4\%}                                  \\ 

\hline 
Frequency (MHz)   & 166.6            & \textbackslash{}                                                       \\
Voltage (MV)      & -1.15            & \textbackslash{}                                                       \\
Phase (rad)       & -0.896           & \textbackslash{}                                                       \\
Bunch length (mm) & \multicolumn{2}{c}{7}                                                                     \\ 
\hline
\multicolumn{3}{c}{double-frequency
  RF system at optimal stretching with bucket height of 4\% (7\%)}        \\ 
\hline
Frequency (MHz)   & 166.6            & 333.2                                                                  \\
Voltage (MV)      & 1.45
  (-1.74)   & -0.504(0.698)                                                          \\
Phase (rad)       & 2.156
  (-0.752) & 2.487(-0.478)                                                          \\
Bunch length (mm) & \multicolumn{2}{c}{43
  (33)}                                                             \\ 
\hline

\multicolumn{3}{c}{double-frequency
  RF system at weak stretching with bucket height of 4\% (solution A)}  $ $  \\ 
\hline
Frequency (MHz)   & 166.6            & 333.2 \\
Voltage (MV)      & -1.46            & 0.528 \\
Phase (rad)       & -0.60            & 0.139 \\
Bunch length (mm) & \multicolumn{2}{c}{16}  \\
\hline
\end{tabular}}}

\end{table}

\begin{figure}[H]
	\centering
	\includegraphics[width=10cm]{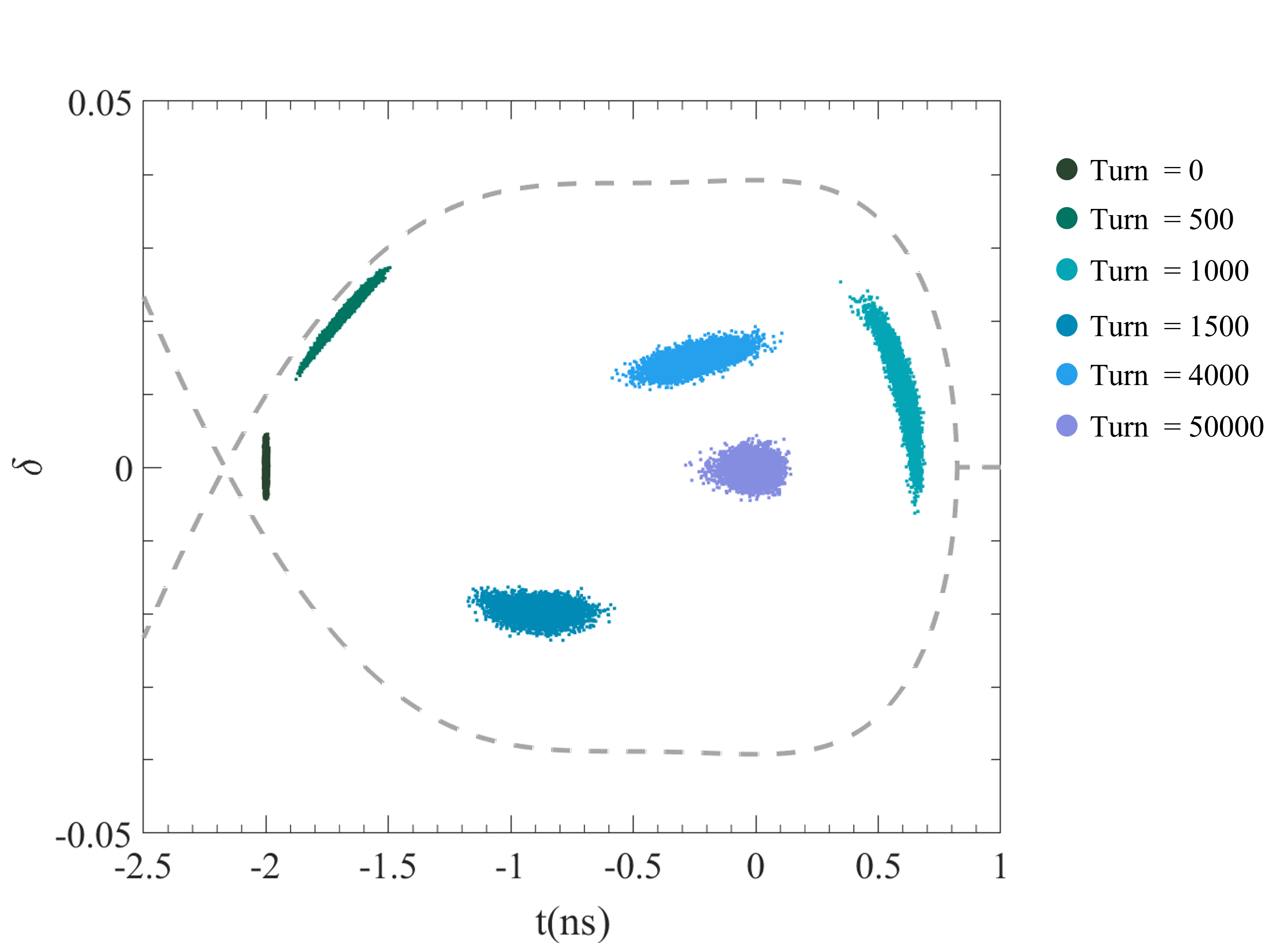}
	\caption{Evolution of the injected bunch in longitudinal phase space over 50000 turns. A double-frequency RF system is used for longitudinal injection, with detailed RF parameters shown in Table 3 (solution A). The bunch is injected at $\delta = 0$ and $t = -2$ ns.}
	\label{Fig4}
\end{figure}

\begin{figure}[H]
	\centering
	\includegraphics[width=8cm]{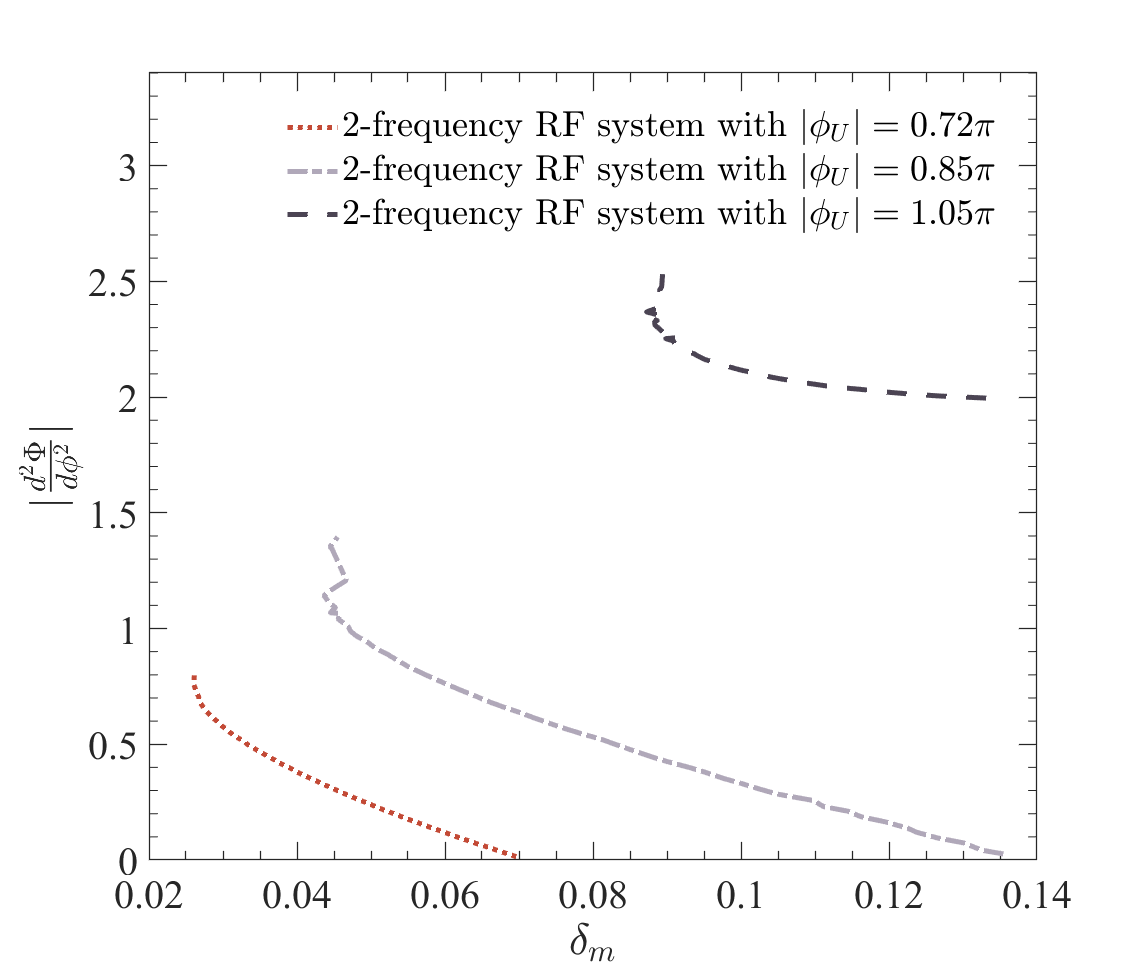}
	\caption{Final front solutions in the objective space for longitudinal injection with a double-frequency RF system and with $|\phi_U|$= 0.72$\pi$, 0.85$\pi$ and 1.05$\pi$.}
	\label{Fig5}
\end{figure}
\subsection{Triple- and quadrupole-frequency RF systems}\label{sec.3.2}
It has been demonstrated in Ref. \cite{jiang_-axis_2018} that with a triple-frequency RF system, it is feasible to realize longitudinal injection in a static bucket while satisfying the optimal bunch lengthening condition. Following the method shown in \cite{jiang_-axis_2018}, we can find a solution for the SAPS with the bucket height $\delta_m$ of 4\%, which has $|\phi_U|$ of 0.9$\pi$, corresponding to $\Delta t$ of 2.7 ns. This solution (denoted as solution B, see Table 4 for detailed parameters) has a lengthening factor of 4.86.

With the proposed optimization method, we can also obtain this solution, i.e., the solution with $(\delta_m, |\frac{d^2\Phi(0)}{d^2\phi}|) = (0.04, 0)$ for $|\phi_U| = 0.9\pi$. Moreover, as shown in Fig. 6, we can find much more candidate solutions, especially those with $|\phi_U|$ larger than 0.9$\pi$. 

In the final front solutions of $|\phi_U| = 0.98\pi$ ($\Delta t = 2.94$ ns), one can find a weak stretching solution (denoted as solution C, parameters also listed in Table 4) with  $(\delta_m, |\frac{d^2\Phi(0)}{d^2\phi}|) = (0.04, 0.13)$ and a lengthening factor of 3.76. Although solution C has a lengthening factor 1.3 times smaller than that of solution B, it allows a larger separation between the injected and stored bunches (see Fig. 7) and promises an increase of the required falling time of the kicker pulse by 0.24 ns.

Furthermore, we test the possibility of achieving $\pi$ phase separation with a triple-frequency RF system. The optimized results are also shown in Fig. 6. For $|\phi_U| = 1.05\pi$, there is a solution with $(\delta_m, |\frac{d^2\Phi(0)}{d^2\phi}|) = (0.04, 0.53)$. The lengthening factor for this weak stretching solution is, however, only about 1.14, implying the potential of bunch lengthening of a triple-frequency RF system is not fully explored. It seems not very attractive to realize a $\pi$-separation with a triple-frequency RF system.

We then make optimization of longitudinal injection with a quadruple-frequency RF system. The optimized results in the objective space for the case of $|\phi_U| = 1.05\pi$ are also shown in Fig. 6 for comparison. In the final front solutions, a solution (denoted as solution D, parameters also listed in Table 4) under the condition of optimal stretching has a bucket height of about 2.6\%, which is much smaller than that with a triple-frequency RF system and less than the MA of the SAPS. We then preform ELEGANT simulation for this solution, where the bunch is injected at $t = -3$ ns and $\delta = 0$. The simulation results are shown in Fig. 8. It indicates that with a quadruple-frequency RF system, it is feasible to realize longitudinal injection in SAPS with $\pi$ phase separation.

\begin{table}
\centering
\caption{RF voltages, phases of RF cavities and equilibrium beam lengths for Solution B, C and D}\label{tb4}
\resizebox{\textwidth}{30mm}{
\setlength{\tabcolsep}{8mm}{
\begin{tabular}[h]{lcccc} 
\hline
\multicolumn{5}{c}{triple-frequency RF system at optimal stretching with $\phi_U = 0.9\pi$ (Solution B)}       \\ 
\hline
Frequency (MHz)   & 166.6  & 333.2  & 499.8  & \textbackslash{}       \\
Voltage (MV)      & 1.640  & -0.908 & 0.343  & \textbackslash{}       \\
Phase (rad)       & 2.738  & -2.423 & -1.433 &  \textbackslash{}       \\
Bunch length (mm) & \multicolumn{4}{c}{34}        \\ 
\hline
\multicolumn{5}{c}{triple-frequency RF system at weak stretching with $\phi_U = 0.98\pi$ (Solution C)}              \\ 
\hline
Frequency (MHz)   & 166.6  & 333.2  & 499.8                                                                                          &  \textbackslash{}       \\
Voltage (MV)      & 1.631  & 0.953  & 0.331                                                                                          &  \textbackslash{}      \\
Phase (rad)       & 2.949  & 1.096  & -0.900                                                                                         &  \textbackslash{}      \\
Bunch length (mm) & \multicolumn{4}{c}{26.3}                                                                                             \\ 
\hline
\multicolumn{5}{c}{quadrupole-frequency RF system at optimal stretching with $\phi_U = 1.05\pi$ (Solution D)}         \\ 
\hline
Frequency (MHz)   & 166.6  & 333.2  & 499.8                                                                                          & 666.4  \\
Voltage (MV)      & -1.64  & -1.12  & 0.55                                                                                           & 0.17   \\
Phase (rad)       & -0.108 & -1.807 & -0.420                                                                                         & -2.21  \\
Bunch length (mm) &  \multicolumn{4}{c}{35}                                                                                                \\
\hline
\end{tabular}}}
\end{table}

\begin{figure}[H]
	\centering
	\includegraphics[width=8cm]{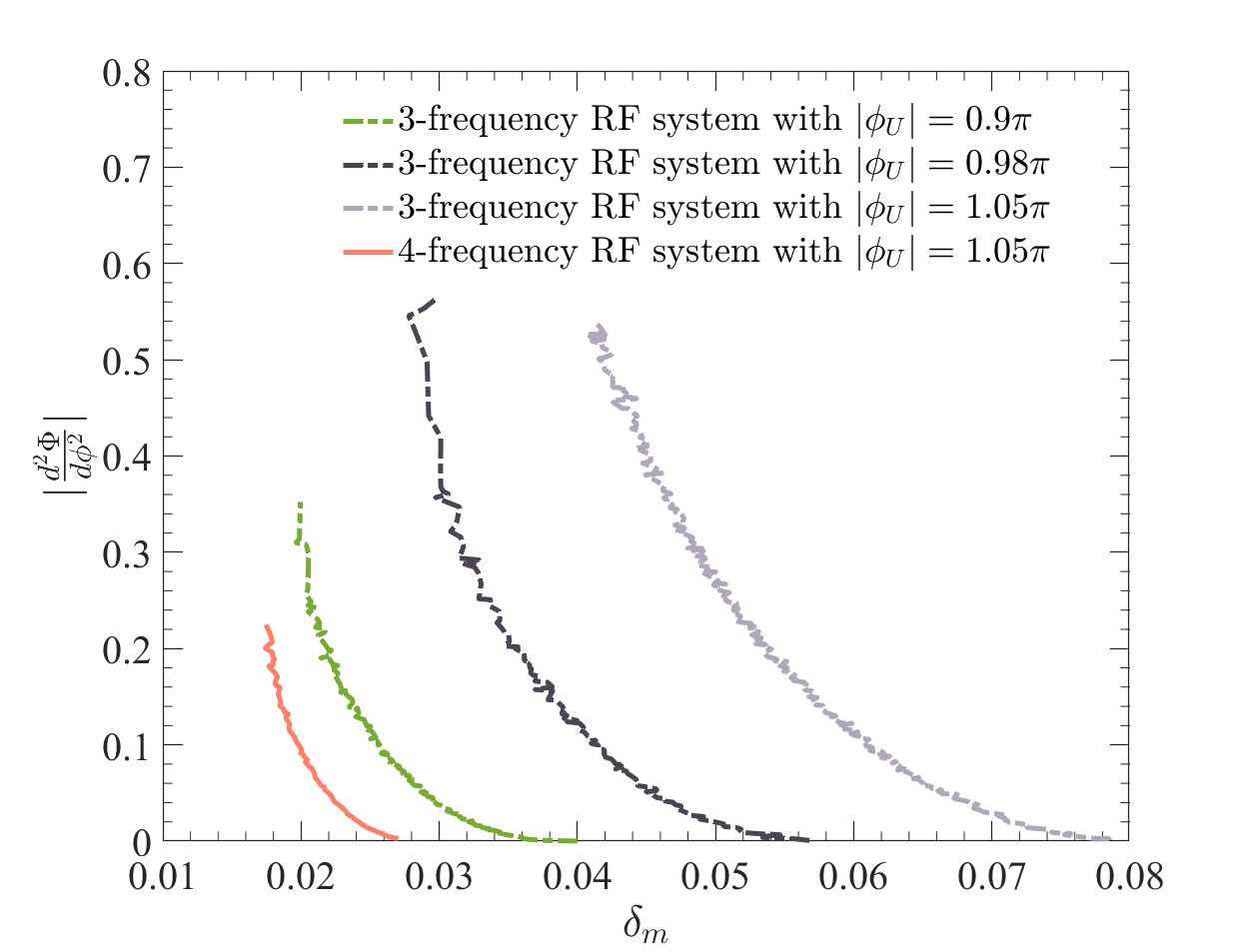}
	\caption{Final front solutions in the objective space for longitudinal injection with triple- and quadrupole-frequency RF systems and with different$|\phi_U|$.}
	\label{Fig6}
\end{figure}

\begin{figure}[H]
	\centering
	\includegraphics[width=12cm]{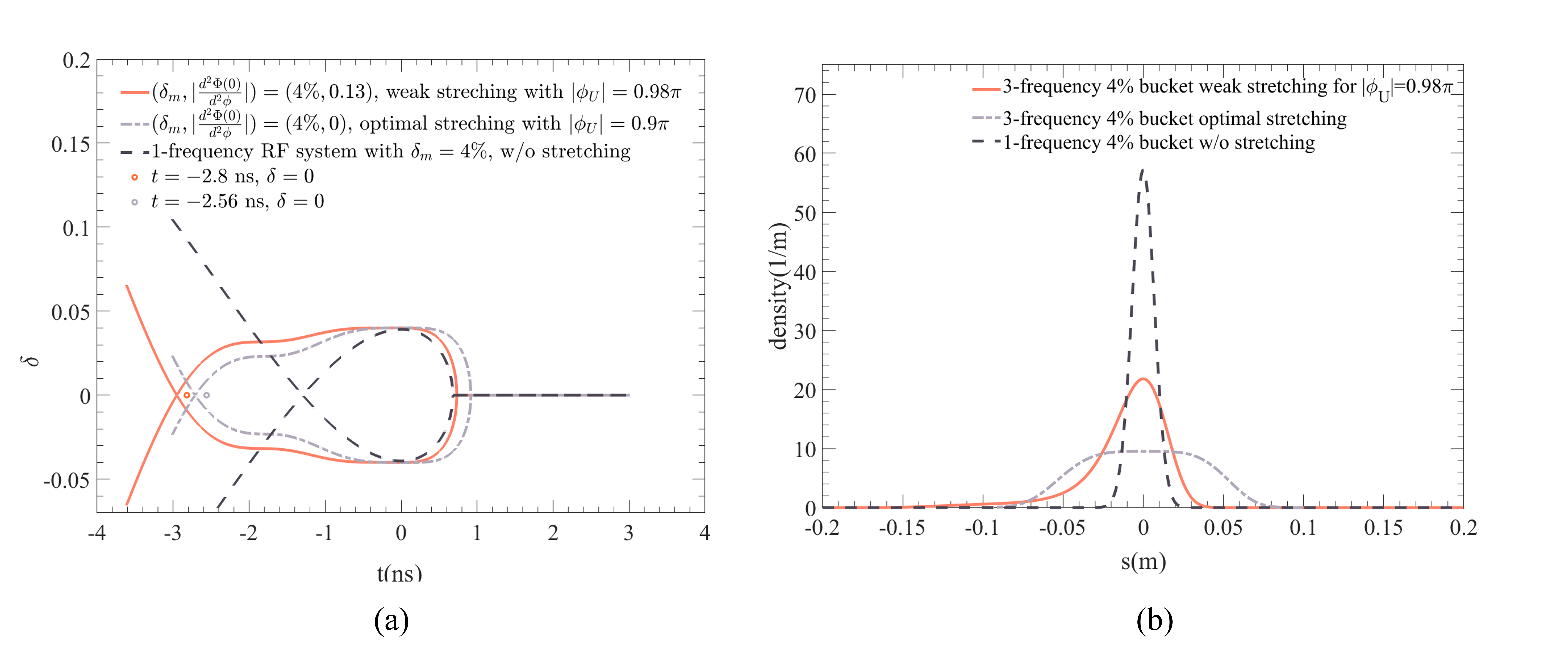}
	\caption{Separatrix (a) and charge density (b) with a triple-frequency RF system and two different RF parameter settings (Solution B and C). These two solutions have the same $\delta_m$ of 4\% but allow different bunch spacing. The separatrix and charge density in presence of a single-frequency RF system is also presented.}
	\label{Fig7}
\end{figure}

\begin{figure}[H]
	\centering
	\includegraphics[width=8cm]{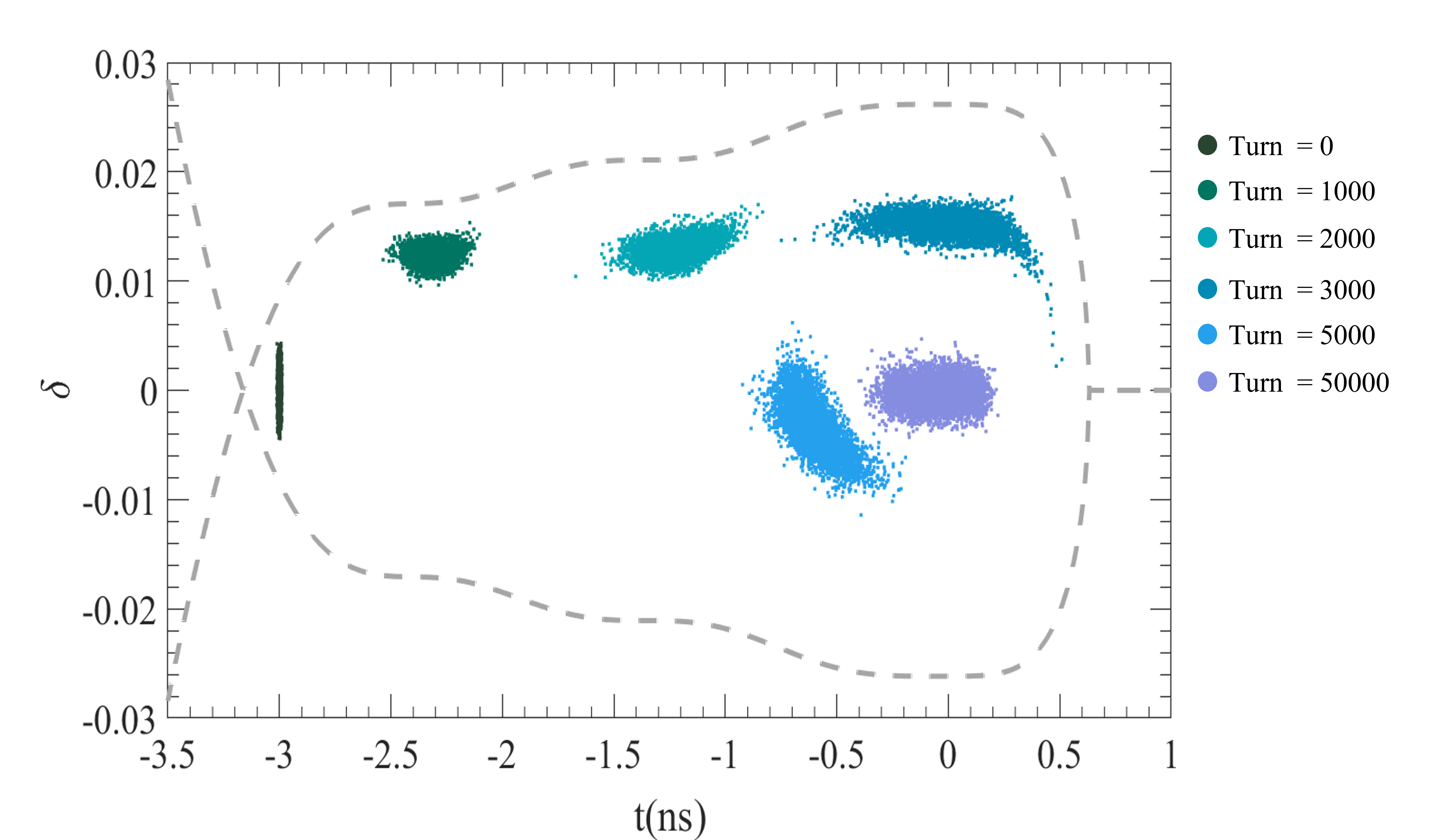}
	\caption{Evolution of the injected bunch in longitudinal phase space over 50000 turns. A quadruple-frequency RF system is used for longitudinal injection, with detailed RF parameters shown in Table 4 (solution D). The bunch is injected at $\delta = 0$ and $t = -3$ ns.}
	\label{Fig8}
\end{figure}

\section{Conclusion}\label{sec.4}
 We have presented a multi-objective optimization method for longitudinal injection with a multi-frequency RF system. By taking the SAPS as an example, we illustrate the application of the proposed method to design the longitudinal injection in a 4GLS. This method enables one to explore a plethora of candidate solutions with different trade-offs among more than one design targets, e.g., a large separation between injected and stored bunches, a large bunch lengthening factor, and a reasonable requirement on the momentum acceptance of the storage ring. From various solutions, one can choose that most adapted to the specific physics design of a 4GLS and the available technical level of the kicker. 

The optimization studies based on the SAPS indicate that in the case of longitudinal injection in a static bucket (namely, with invariant RF parameters during injection), introducing harmonic cavities of more RF frequencies is an effective way to increase the time separation between injected and stored bunches. Of course, in such approach, one must pay some prices, e.g., more complex RF system and higher cost.
 
Especially, the studies show that it would be possible to realize longitudinal injection in a static bucket with only a double-frequency RF system, if not taking the optimal bunch lengthening as a condition that must be satisfied. Considering that double-frequency RF systems are commonly adopted in 4GLSs for bunch lengthening and mitigation of collective effects, this result suggests a cost-efficient and easy-to-apply longitudinal injection option for 4GLSs. More design and experimental efforts are expected to turn it into a reality.

\section*{Acknowledgments}
This work was supported by the Guangdong Basic and Applied Basic Research Foundation (2019B1515120069), National Natural Science Foundation of China (No. 11922512), Youth Innovation Promotion Association (No. Y201904), and Bureau of Frontier Sciences and Education of Chinese Academy of Sciences (No. QYZDJ-SSW-SLH001).

\section*{Appendix. Potential function of a multi-frequency RF system}
The derivation of a potential function for an $N$-frequency RF system is straightforward by using the following form of a normalized voltage function (normalized to $U_0$)
\begin{equation}
\label{eqA.1}\\
V=\sum_{i=1}^{N} v_{i} \sin \left(n_{i} \phi+\phi_{i}\right).\tag{A. 1}
\end{equation}
The potential function $\Phi$ is relative to the equation of motion through
\begin{equation}
\label{eqA.2}\\
\frac{d \Phi}{d \phi}=-\frac{E_{0} T_{0}}{U_{0}} \frac{d \delta}{d t}=1-V.\tag{A. 2}
\end{equation}
The coefficient in front of $\frac{d\delta}{dt}$ is due to the normalized definition of $\Phi$ and $V$. 

From Eq.\eqref{eqA.1} and \eqref{eqA.2}, the potential function can be solved as

\begin{equation}
\label{eqA.3}\\
\Phi=\phi+\sum_{i=1}^{N} \frac{v_{i}}{n_{i}} \cos \left(n_{i} \phi+\phi_{i}\right)+C,\tag{A. 3}
\end{equation}
where $C$ is an arbitrary constant. 

For convenience, we take this constant to be
\begin{equation}
\label{eqA.4}\\
C=-\sum_{i=1}^{N} \frac{v_{i}}{n_{i}} \cos \left(\phi_{i}\right),\tag{A. 4}
\end{equation}
which will let $\Phi = 0$ when $\phi$ is zero. 







\section*{Reference}
\biboptions{sort&compress}
\bibliographystyle{elsarticle-num}

\end{document}